\documentclass[a4paper,11pt]{article}

\usepackage{contribution}

\newcommand{\weblink}[2][]{%
    \ifthenelse{\equal{#1}{}}%
    {\textnormal{\url{#2}}}%
    {\textnormal{\href{#2}{#1}}}%
}

\def\beq{\begin{equation}}
\def\eeq#1{\label{#1}\end{equation}}
\def\eeqn{\end{equation}}

\def\beqa{\begin{eqnarray}}
\def\eeqa#1{\label{#1}\end{eqnarray}}
\def\eeqan{\end{eqnarray}}

\let\bar=\overbar

\def\Dslash{\not{\hbox{\kern-4pt $D$}}}
\def\dslash{\not{\hbox{\kern-2pt $\del$}}}

\def\msb{{\bar{\ssstyle M \kern -1pt S}}}

\newcommand{\contribution}[7][]{%
  \clearpage
  \thispagestyle{plain}
  \ifthenelse{\equal{#1}{}}
  {\hypersetup{pdftitle={#2}}}
  {\hypersetup{pdftitle={#1}}}
  \hypersetup{pdfauthor={{#3} {#4}}}
  {\centering\normalfont\LARGE\bfseries\sffamily #2 \par\nobreak}
  \lhead{}
  \chead{%
    \textit{\footnotesize XIV International Conference on Hadron Spectroscopy
      (\weblink[\textit{hadron2011}]{http://www.hadron2011.de}), 13-17 June 2011, Munich, Germany}%
  }
  \rhead{}
  \bigskip
  \begin{center}
    {#3} {#4}\ifthenelse{\equal{#6}{}}{}{\footnote{\weblink[#6]{mailto:#6}}}
    \ifthenelse{\equal{#7}{}}{}{#7} \\
    \textit{#5}
  \end{center}
  \bigskip
}

\renewcommand{\abstract}[1]{%
  \begin{center}
    \begin{minipage}{0.85\textwidth}
      \begin{footnotesize}
        #1
      \end{footnotesize}
    \end{minipage}
  \end{center}
  \bigskip
}

\begin{document}

%
%
%
%
%
{  


%
\makeatletter
\@ifundefined{c@affiliation}%
{\newcounter{affiliation}}{}%
\makeatother
\newcommand{\affiliation}[2][]{\setcounter{affiliation}{#2}%
  \ensuremath{{^{\alph{affiliation}}}\text{#1}}}

\contribution[Hadron Resonances] {Hadron Resonances Within a
Constituent-Quark Model}
{Regina}{Kleinhappel}  
{\affiliation[Physics Institute, University of Graz]{1} \\
} {regina.kleinhappel@uni-graz.at} {\!\!$^,\affiliation{1}$ and
Wolfgang Schweiger\affiliation{1}}

%

\abstract{%
In order to get a more realistic description of the hadron spectrum
we extend a constituent-quark model by explicit mesonic degrees of
freedom. The resulting system of constituent (anti)quarks, which are
subject to an instantaneous confining force, and mesons, which
couple directly to the quarks, is treated by means of a relativistic
coupled-channel framework. It can be formally shown that the
mass-eigenvalue problem for such a system is equivalent to a
hadronic eigenvalue problem in which the eigenstates of the pure
confinement potential (bare hadrons) are coupled via meson loops.
Following this kind of approach we have calculated hadron masses and
decay widths for a simple toy model.
                                                                               }
%

The resonance character of hadron excitations is usually not taken
into account in constituent quark models. As a consequence most of
the (perturbatively) calculated partial decay widths come out too
small as compared to experiment \cite{Melde:2006yw}. This suggests
that physical hadron resonances are not just simple bound states of
valence (anti)quarks, but should also contain higher Fock
components. We propose to model the additional quark-antiquark pairs
by means of mesons which can couple directly to the valence
(anti)quarks.

A natural starting point for this kind of description is the chiral
constituent quark model \cite{Glozman:1995fu}. Within this model it
is assumed that the effective degrees of freedom emerging from the
spontaneous breaking of chiral symmetry are (confined) constituent
(anti)quarks and Goldstone bosons, i.e. the lightest pseudoscalar
mesons. We use the point form of relativistic quantum mechanics in
connection with the Bakamjian-Thomas construction to calculate mass
spectra and decay widths. This kind of approach is Poincar\'e
invariant and one only has to deal with an eigenvalue problem for an
appropriately defined mass operator~\cite{Biernat:2010tp}.

In order to allow for the decay of hadron excitations into a lower
lying state by emission of a Goldstone boson we adopt a 2-channel
mass operator. A general mass eigenstate has then a
valence-(anti)quark component $\vert \psi_{\mathrm{val}}\rangle$ and
a valence-(anti)quark + Goldstone-boson component $\vert
\psi_{\mathrm{val+GB}}\rangle$. These components are coupled via
vertex operators $\hat{K}$ and $\hat{K}^\dag$ that describe the
emission and absorption of the Goldstone boson by the (anti)quark,
respectively. If $\vert \psi_{\mathrm{val+GB}}\rangle$ is eliminated
by means of a Feshbach reduction one ends up with a mass-eigenvalue
equation for $\vert \psi_{\mathrm{val}}\rangle$ which takes on the
form:
\begin{equation}\label{eq:mev}
\big(\hat{M}_{\mathrm{val}}+\underbrace{\hat K^\dagger (m-\hat
M_{\mathrm{val}+GB}+i0)^{-1}\hat
K}_{\hat{V}_{\mathrm{opt}}(m)}\big)\vert \psi_{\mathrm{val}}\rangle
= m \vert \psi_{\mathrm{val}}\rangle \, .
\end{equation}

\vspace{-0.5cm}\noindent The channel mass operators
$\hat{M}_{\mathrm{val}}$ and $\hat{M}_{\mathrm{val}+GB}$ consist of
a kinetic-energy term and an instantaneous confining potential. By
expanding $\vert\psi_{\mathrm{val}}\rangle$ in terms of eigenstates
of $\hat{M}_{\mathrm{val}}$ (which we call \lq\lq bare hadrons\rq\rq)
Eq.~(\ref{eq:mev}) can be converted into a system of algebraic equations for
the expansion coefficients. Physically speaking, this system represents
again a (non-linear) mass-eigenvalue problem, but on the hadronic level rather
than on the quark level. It describes the coupling of bare hadrons
via Goldstone-boson loops and has to be solved self-consistently (for details, see
Ref.~\cite{Kleinhappel:2010}).

As a first test we have applied these ideas to a simple toy model in
which spin and isospin are completely neglected and the bare hadrons
are just quark-antiquark pairs confined by a harmonic oscillator
potential. In order to give the model some physical meaning we have
adjusted the parameters such that the masses of the ground state and
the first excited state of the $\omega$ are approximately
reproduced. As can be seen in Fig.~\ref{DepCoCo} the decay width of
the first excited state exhibits a maximum of about $26$~MeV as a
function of the Goldstone-boson-(anti)quark coupling strength and
vanishes as soon as the real part of the mass eigenvalue approaches
the lowest threshold. We observe a considerable increase
of the decay width as compared to perturbative calculations. This
gives some hope that typical decay widths of $0.1$~GeV or even more
can be achieved for baryon resonances within the full chiral
constituent-quark model.
\begin{figure}
\centering
\includegraphics[width=0.65\textwidth]{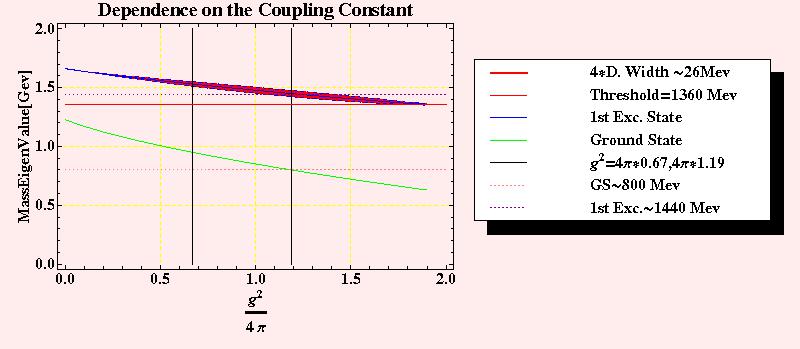}
\vspace{-0.3cm}\caption{\small{Predictions of our toy model. The
mass of the ground state (green line) and the first excited state
(blue line) as functions of the Goldstone-boson-quark coupling. The
red band between the dashed blue lines indicates the decay width of
the first excited state (multiplied by a factor 4 for better
visibility). The range of couplings allowed by the
Goldberger-Treiman relation is indicated by the black vertical
lines.}} \label{DepCoCo}
\end{figure}
%

\noindent{\bf Acknowledgement:} R. Kleinhappel acknowledges the
support of the \lq\lq Fonds zur F\"orderung der wissenschaftlichen
Forschung in \"Osterreich\rq\rq\ (FWF DK W1203-N16)

\vspace{-0.5cm}

%

}  


\end{document}